\documentclass[12pt]{article}
\usepackage{units}
\usepackage{cite,mcite}
\usepackage{graphicx}
\usepackage{subfigure}
\usepackage{mathptmx}
\usepackage{helvet}
\usepackage{amsmath}

\def\be{\begin{equation}}
\def\ee{\end{equation}}
\newcommand{\beq}{\begin{equation}}
\newcommand{\eeq}{\end{equation}}
\newcommand{\ba}{\begin{eqnarray}}
\newcommand{\ea}{\nonumber \end{eqnarray}}
\def\mr{\langle r \rangle}

\begin{document}

\vspace*{0.5cm}
\begin{center}
{\Large \bf  Bose-Einsten and other two particle correlations - MC study }\\

\vspace*{1cm}

V.A. Schegelsky \footnote{e-mail addresses: valery.schegelsky@cern.ch}\\

\vspace*{0.5cm}
 Petersburg Nuclear Physics Institute, NRC Kurchatov
 Institute, Gatchina, St.~Petersburg, 188300, Russia \\

\vspace*{1cm} \

\end{center}

\begin{abstract}

    The MC toy model for Bose-Einstein correlations (BEC)  is considered to  make  a best choice from different
    reference distributions. It occurs that the minimal bias in the BEC parameters estimation is provided by the reference sample
    which is being
    simulated from "observed" sample by the turn of all $\vec{p_t}$ vectors in an event by a random $\delta \phi$.

    It proposed to use similar approach in the analysis  $C_2(\Delta \eta, \Delta \phi$) correlation.
    To make the analysis less
    dependent on reference sample, new definition of  $\Delta \phi$ is implemented as the angle between vectors
    $p_1$ and $p_2$ in the transverse
    plane. Then $C_1(\Delta \phi)$ correlation can be studied  without a reference sample subtraction.

    The ridge in
    $C_2(\Delta \eta, \Delta \phi$) correlation  appears if an appropriate cut in Q-value ($Q=p_2-p_1$) is applied.

\end{abstract}

\section{Introduction}
\label{bose}

One of the way to study the space-time picture of hadronic interaction is to measure the correlations between two
identical particles which momenta are close to each other.
 This correlations  are used to estimate the
size of the domain, from which particles produced  are emitted.\\

Consider the amplitude $M(r_1,r_2)$ in coordinate representation in which two identical pions, $\pi_1$ and $\pi_2$,
are emitted at the points $r_1$ and $r_2$ correspondingly.
   The coordinates $r_1$, $r_2$ can not be  measured, and instead one does estimate pions momenta quantities
  $p_1$ and $p_2$ . So we have to take the Fourier transform
\be
M_a(p_1,p_2)=\int \frac{d^4 r_1}{(2\pi)^4}e^{ip_1r_1}\frac{d^4 r_1}{(2\pi)^4}e^{ip_2r_2}M(r_1,r_2).
\ee
Besides this we have to consider the permutation of two identical pions. That is we have to add to $M_a$ the
amplitude
\be
M_b(p_1,p_2)=\int \frac{d^4 r_1}{(2\pi)^4}e^{ip_2r_1}\frac{d^4 r_1}{(2\pi)^4}e^{ip_1r_2}M(r_1,r_2),
\ee
where the pion with momentum $p_2$ was emitted from the point $r_1$ and wise versa. This can be written as
\be
M(p_1,p_2)=M_a+M_b=M_a\cdot (1+e^{irQ}),
\ee
where the 4-vectors $r=r_1-r_2$ and $Q=p_2-p_1$ \footnote{Everywhere we suppose that particles are pions} .

Finally the cross section takes the form
\be
\frac{E_1E_2d^2\sigma}{d^3p_1d^3p_2}=\frac 1{2!}|M_a|^2<2+2e^{irQ}>= |M_a|^2(1+<e^{irQ}>).
\ee

 Here the factor $1/2!$ reflects the identity of two pions and the angular brackets indicates the averaging over the
 ($r_1,r_2$)
space
 distribution. Assuming, for simplicity, the Gaussian form we get
$<e^{irQ}>=e^{-<r^2>Q^2}$.

Thus the width of a peak at small momenta difference $Q\to 0$ characterizes the size $<r>$ of the domain from which
the pions were emitted.

The general definition of a correlation function of 2 variables - $p_1$ and $p_2$ can be written as
\be
C_{2}(p_1,p_2)=\frac {P(p_1,p_2)}{P_{ref}(p_1,p_2)}.
\label{C2}
\ee
The value $P_{ref}(p_1,p_2)$ is called  a reference function, artificial function with switched off the correlation
under search . Usual task is to extract correlation $C_{2}$ from the "measured" value $P(p_1,p_2)$. In the following
we will consider Bose-Einstein correlations (BEC)  in the multi-particle final state (MPS) in pp-collisions. Suppose
that BEC correlation is determined by  $P(p_1,p_2)$ with the reference $P_{ref}(p_1,p_2)$. Usually MC generators
simulate MPS without BEC. In the following the ATLAS MC \cite{atlasmc} of the pp-interaction with 7 TeV center of
mass energy is used. In the MC event sample there is no BE correlations , but known hadronic resonance states are
"produced".

The distributions of identical bosons should be influenced by $BEC$. The proximity in phase space between final state
particles with 4-momenta $p_1$ and $p_2$ can be quantified by
\be
Q = \sqrt{-(p_1 - p_2)^2}.
\label{Q}
\ee
The BEC effect is observed as an enhancement at low $Q\sim 300MeV$. To extract  the effect one can compare measured
Q-spectra with similar one but without BEC, with reference spectra .  Then the ratio
\begin{equation}\label{eq:ratio}
  R(Q) = \frac{\frac{dN}{dQ}-\frac{dN_{ref}}{dQ}}{\frac{dN_{ref}}{dQ}}
\end{equation}

can be fitted with an appropriate formulae

\be \label{eq:fFit}
          R(Q) = F(rQ)+a+bQ
\ee

F(rQ) is the Fourier transform of the spatial distribution of the emission region with an effective size $\mr$. The
function can be parameterize as a linear exponent  $F=Ce^{-r_{e}Q}$ or  a Gaussian $F=Ce^{-r_{g}^{2}Q^{2}}$. The
parameter $b$ may be accounted for a long range correlation.For the same $\mr$ there is quite simple relation between
these two cases :
\be \label{eq:rerg}
         r_{e} = \sqrt{\pi}r_{g}.
\ee
These two choices are connected  to different  radiation sources distribution inside the emission region. The
exponential form approximates the case when radiation is originated from the surface of the sphere with radius $r_e$.
The second form describes the radiation in space  $\sim e^{-r_{g}^{2}Q^{2}}$. The fit of the same distribution
$\frac{dN_{ss}}{dQ}$ with these two cases gives r-parameter ratio as in (~\ref{eq:rerg}). The exponential fit is a
bit  better and will be used in the following.

\section{ MC BEC model and different reference functions}

    Two values of $\frac{dN}{dQ}$ are measured in the experiment:  $\frac{dN_{ss}}{dQ}$ for the like sign particle pairs
    and $\frac{dN_{us}}{dQ}$ for the unlike sign case.
     To simulate BEC into MC distribution , each entry into $\frac{dN_{ss}}{dQ}$   can be
    weighted by  the function
\be
         W(Q) = 1.+ e^{-rQ}
\ee
\label{W(Q)}
to produce $\frac{dN_{mod}}{dQ}$. The value  $r$ is fixed equal to 2fm.
   Of course, such model is the oversimplified  toy but good enough to illustrate difference between reference functions.

\begin{figure}
\begin{center}

\subfigure[The  difference of normalized distributions  $ \frac{dN_{mod}}{dQ}- \frac{dN{_{ss}}^{ref}}{dQ}$ ]{
\includegraphics[width=.5\textwidth]{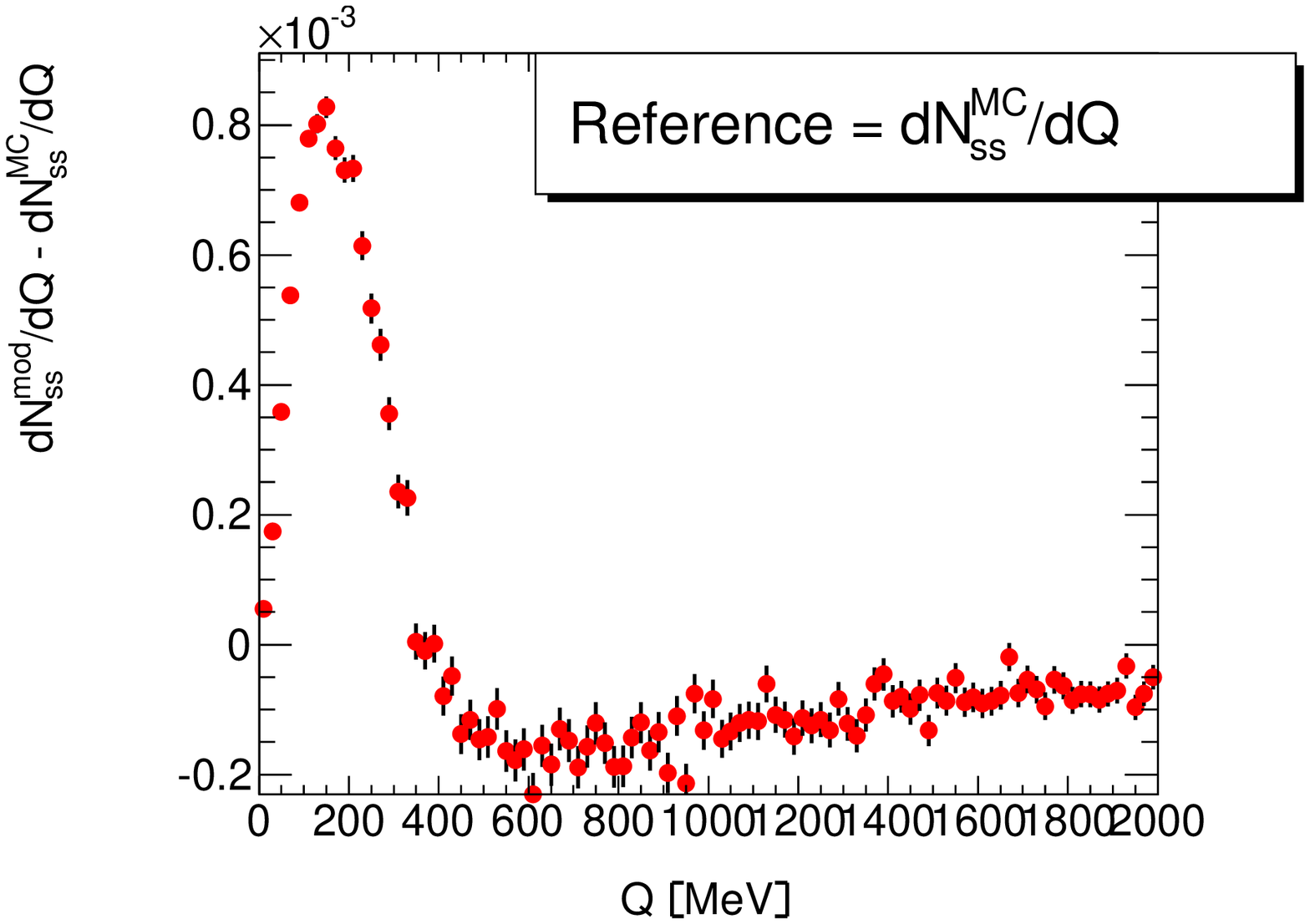}
\label{fig:dif_mod_mc}
}
\subfigure[Fit of the ratio
$R(Q)=\frac{\frac{dN_{mod}}{dQ}-\frac{dN{_{ss}}^{ref}}{dQ}}{\frac{dN{_{ss}}^{ref}}{dQ}}$]{
\includegraphics[width=.5\textwidth]{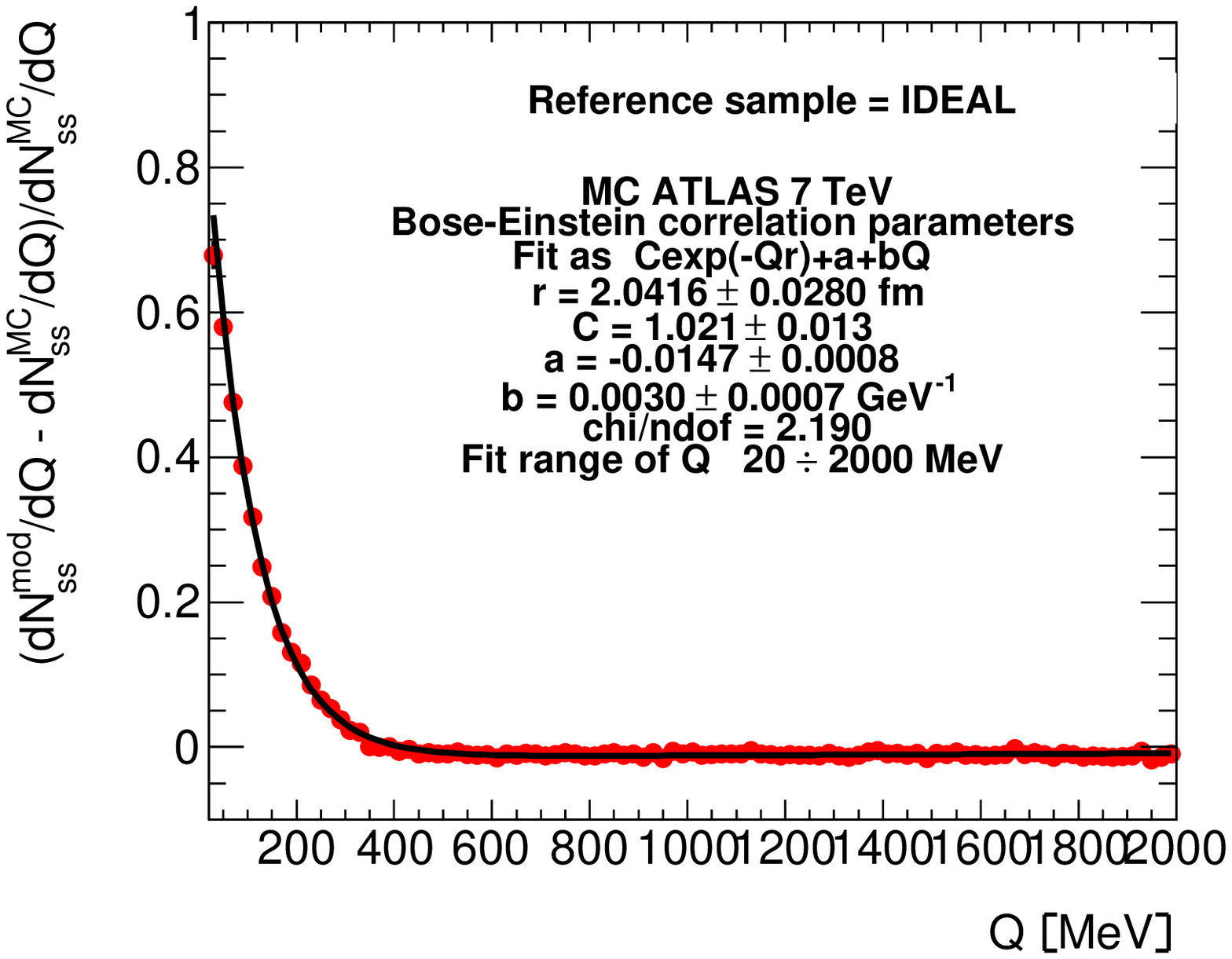}
\label{fig:fit_mod_mc}
}
\end{center}
    \caption{ MC Model fit with the "ideal" reference}
     \label{fig:ideal_case}
\end{figure}

  First consider the ideal case when MC $\frac{dN_{ss}}{dQ}$ will be used as the reference function
  $\frac{dN^{ref}_{ss}}{dQ}$. Fig ~\ref{fig:ideal_case} shows the difference $\frac{dN_{mod}}{dQ}-\frac{dN^{ref}_{ss}}{dQ}$ and the result
  of the fit of the ratio (\ref{eq:ratio})
with the formulae (\ref{eq:fFit})
\footnote { If the  parameter $a$ or $b$ is equal zero within two
  statistical errors it is fixed to be zero. The estimations of parameters error are "normalized" to make $\chi^2/Ndof =1$ }.
  The result is somewhat trivial and can be considered as a "goodness" of the MC
  generator. The procedure used might be looked as switching ON the BEC correlation, to add such correlation to MC
  procedure.

  In the BEC analyses of experimental data different approaches have been explored and it is worth to compare
  known reference functions with the use of the toy MC model.

\begin{figure}
\begin{center}
   \subfigure[The  difference of normalized distributions  ($ \frac{dN_{mod}}{dQ}- \frac{dN_{us}}{dQ}$ )]{
 \includegraphics[width=.5\textwidth]
{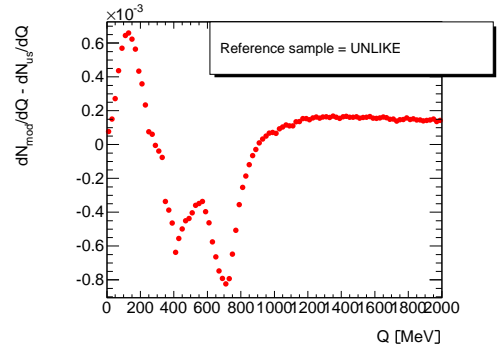}
\label{fig:dif_mod_unli}
}
 \subfigure[Fit of the ratio $ R(Q) = \frac{\frac{dN_mod}{dQ}- \frac{dN_{us}}{dQ}}{\frac{dN_us}{dQ}}$]{

\includegraphics[width=.5\textwidth]
{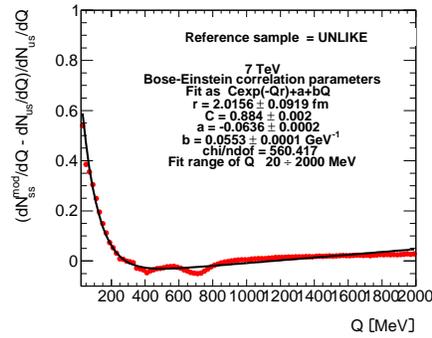}
\label{fig:fit_mod_unli}
}
\end{center}
    \caption{ MC Model fit with the $\frac{dN}{dQ}$ for unlike sign pairs as the reference}
     \label{fig:unlike_case}
\end{figure}

  The unlike Q-distribution looks as a natural choice because there should be no BEC for the unlike sign pairs.
   The source radius from the fit  is close to
  the model value  ( Fig ~\ref{fig:unlike_case}), however the resonances contribution makes fit quality bad. The
  contributions of the $\rho^0$ and the remnants of $\eta$ mesons are clearly seen.

 One has to find a way to switch OFF BEC from the "observed"  $\frac{dN_{mod}}{dQ}$  distributions.
 It looks as an easiest way  to construct the   $\frac{dN{_{ss}}^{ref}}{dQ}$ from two independently detected events:
  one momentum vector is taken from
 event under analysis, and the second - from a preceding events. The result is strongly dependent on
 requirements of such event  selection. Moreover, such requirements formulation is rather arbitrary.

  Another approach is to emulate independent event sample with the use of the  events set under consideration.
   The "referenced" event will
   have at least the same particle multiplicity,the same numbers
 of positive and negative charged particles.
  Evidently, such procedure will work as a blind correlation terminator because there
 is no any BEC "marker" on a particular entry. The hope can be that pairs with small Q-values will be replaced by
 pairs which had higher Q-value before the transformation.  The  reference distribution
  $\frac{dN{_{ss}}^{ref}}{dQ}$ will contain one
 vector from the real event and another from the reference one.

\begin{figure}
\begin{center}
 \subfigure[The  difference of normalized distributions  $ \frac{dN_{mod}}{dQ}-\frac{dN{_{ss}}^{ref}}{dQ}$ ]{
\includegraphics[width=.5\textwidth]
{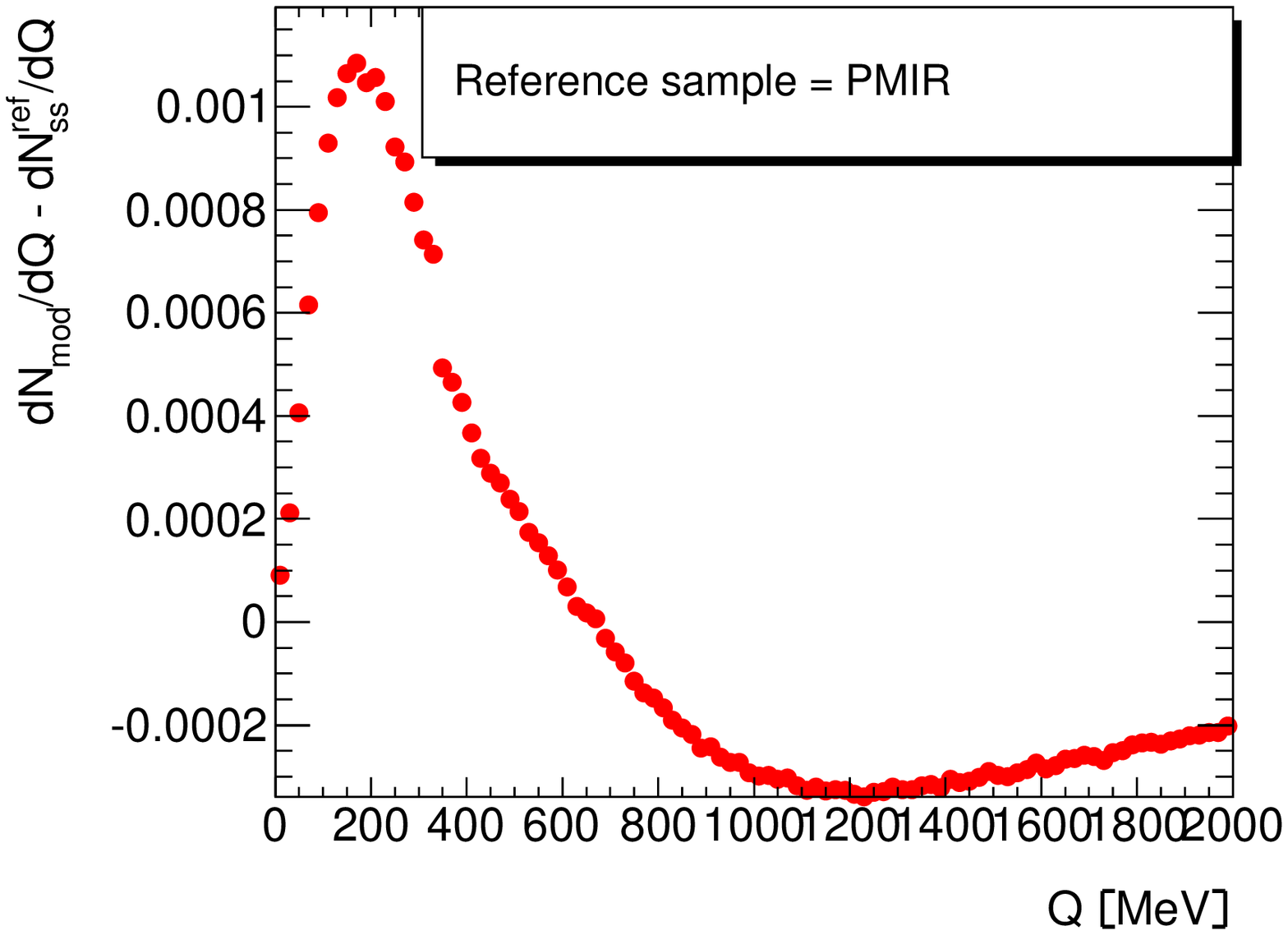}
\label{fig:dif_mod_pmir}
}
 \subfigure[Fit of the ratio $ R(Q) = \frac{\frac{dN_mod}{dQ}- \frac{dN{_{ss}}^{ref}}{dQ}}{\frac{dN{_{ss}}^{ref}}{dQ}}$ ]{
\includegraphics[width=.5\textwidth]
{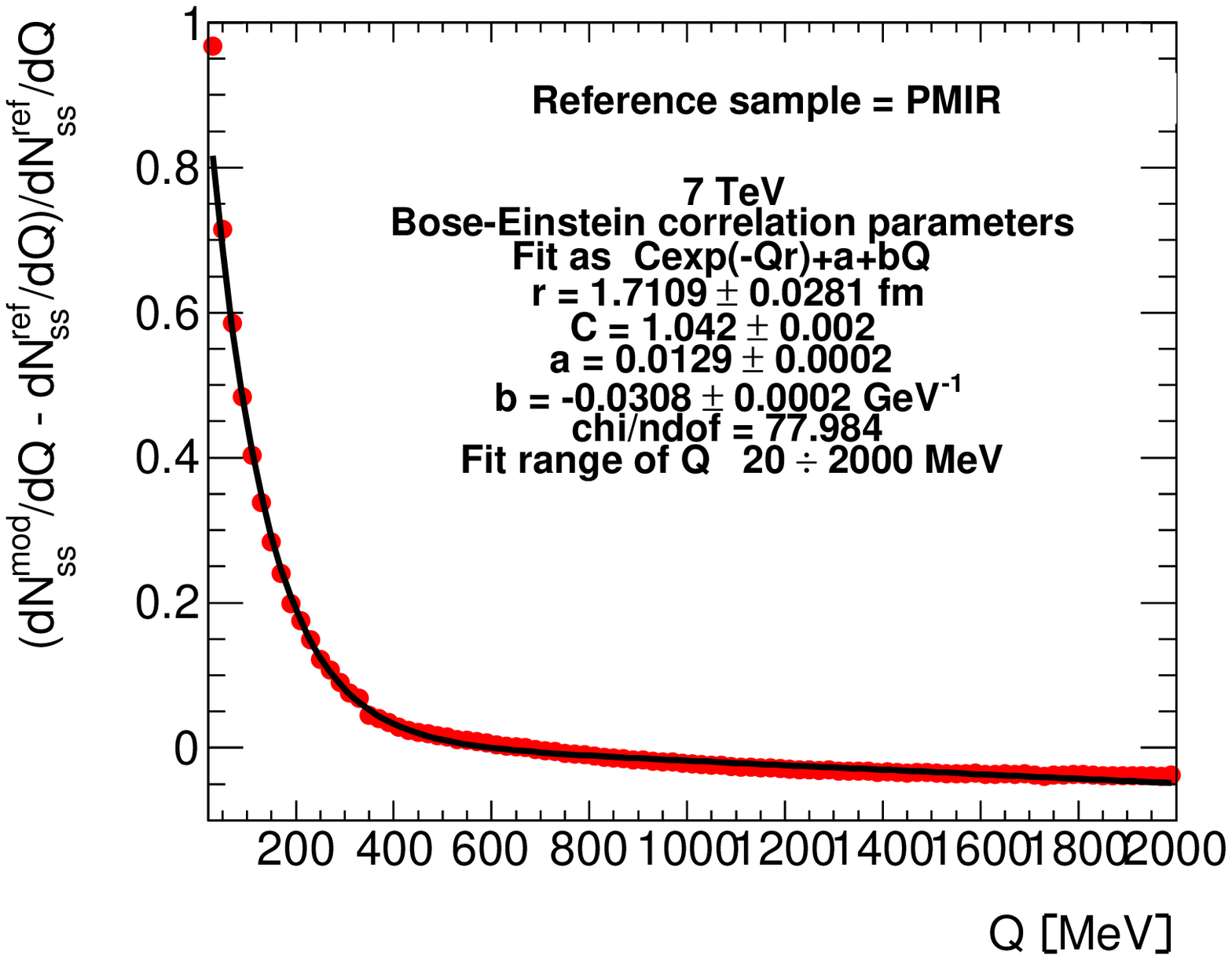}
\label{fig:fit_mod_pmir}
}
\end{center}
    \caption{ MC Model fit of the ratio $ R(Q)$ where the reference sample consists of real events  with all 3-vector  momentum reflected in space  }
     \label{fig:pmir_case}
\end{figure}

  The standard procedure to prepare mirror sample
   from the
 original  one is to change in the event all momenta 3-vectors as  $ \vec p \rightarrow - \vec p $ (PMIR case).
 It would be good if events
 (and the detector) have such
  symmetry. The reference sample is the result of measurements with the detector identical to the real one but
  in the mirror space. Fig ~\ref{fig:pmir_case}  illustrate results of such
 procedure. We do see that  the distribution  $\frac{dN{_{ss}}^{ref}}{dQ}$ is  different from the model too much. In particular
 there is no entries in the reference distribution if Q-value in the real events is
 smaller than $\sim 300 MeV$ : the procedure  removes all small Q entries  from the reference Q-distribution,
 not only BEC entries.
 In other words,
  the BEC contribution will be overestimated and
 the  ratio  (\ref{eq:fFit}) Q-distribution will be wider ( compared with  Fig ~\ref{fig:ideal_case}) and the $r$-parameter  of
 (\ref{eq:fFit})  will be smaller  than  in the model.

\begin{figure}
\begin{center}
 \subfigure[The  difference of normalized distributions  $ \frac{dN_{mod}}{dQ}-\frac{dN{_{ss}}^{ref}}{dQ}$ ]{
\includegraphics[width=.5\textwidth]
{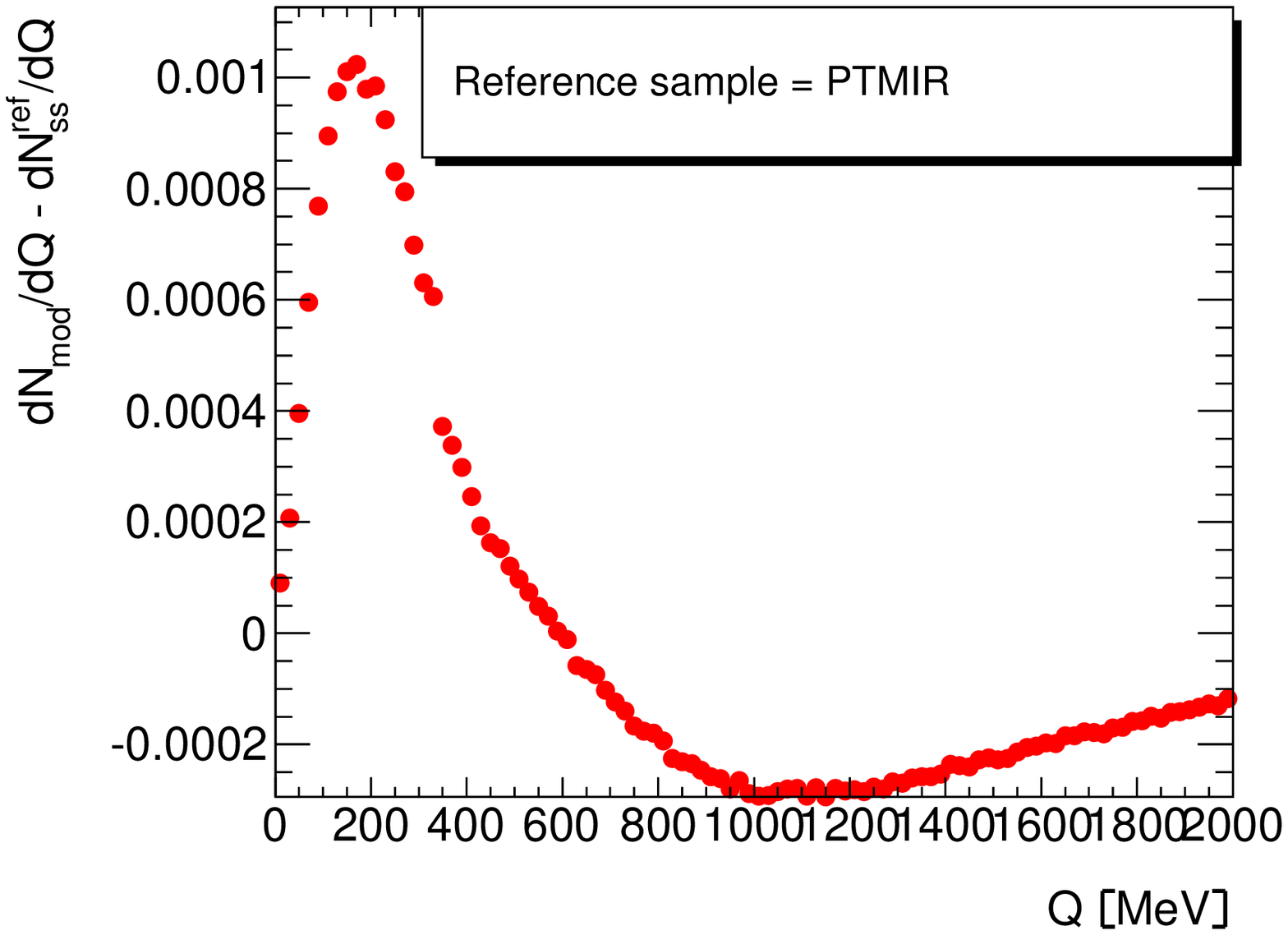}
\label{fig:dif_mod_ptmir}
}
 \subfigure[Fit of the ratio $ R(Q) = \frac{\frac{dN_mod}{dQ}- \frac{dN{_{ss}}^{ref}}{dQ}}{\frac{dN{_{ss}}^{ref}}{dQ}}$ ]{
\includegraphics[width=.5\textwidth]
{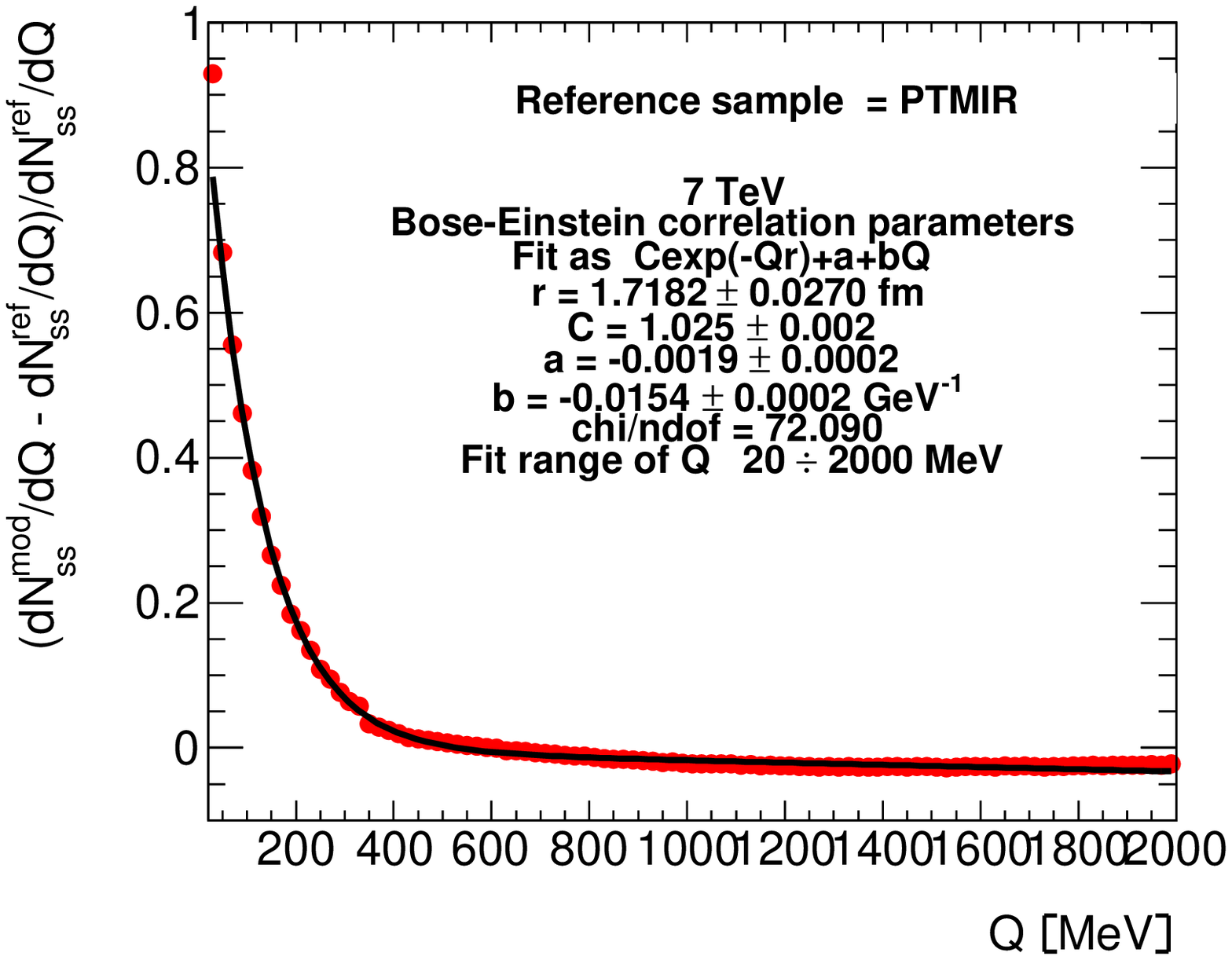}
\label{fig:fit_mod_ptmir}
}
\end{center}
    \caption{  MC Model fit of the ratio $ R(Q)$ where the reference sample consists of real events  with all $\vec p_t$
momentum reflected in the transverse plane  }
     \label{fig:ptmir_case}
\end{figure}

  The better way might  be if the reflection is made only in the transverse plane
  $\vec p_t \rightarrow - \vec p_t$ (PTMIR case). The results are a bit closer to the model values but still
  there is no small Q
  events in the reference distribution if the model Q-values is smaller than $\sim 300 MeV$ ( Fig \ref{fig:ptmir_case}).

\begin{figure}
\begin{center}
 \subfigure[The  difference of normalized distributions  $ \frac{dN_{mod}}{dQ}-\frac{dN{_{ss}}^{ref}}{dQ}$ ]{
\includegraphics[width=.5\textwidth]
{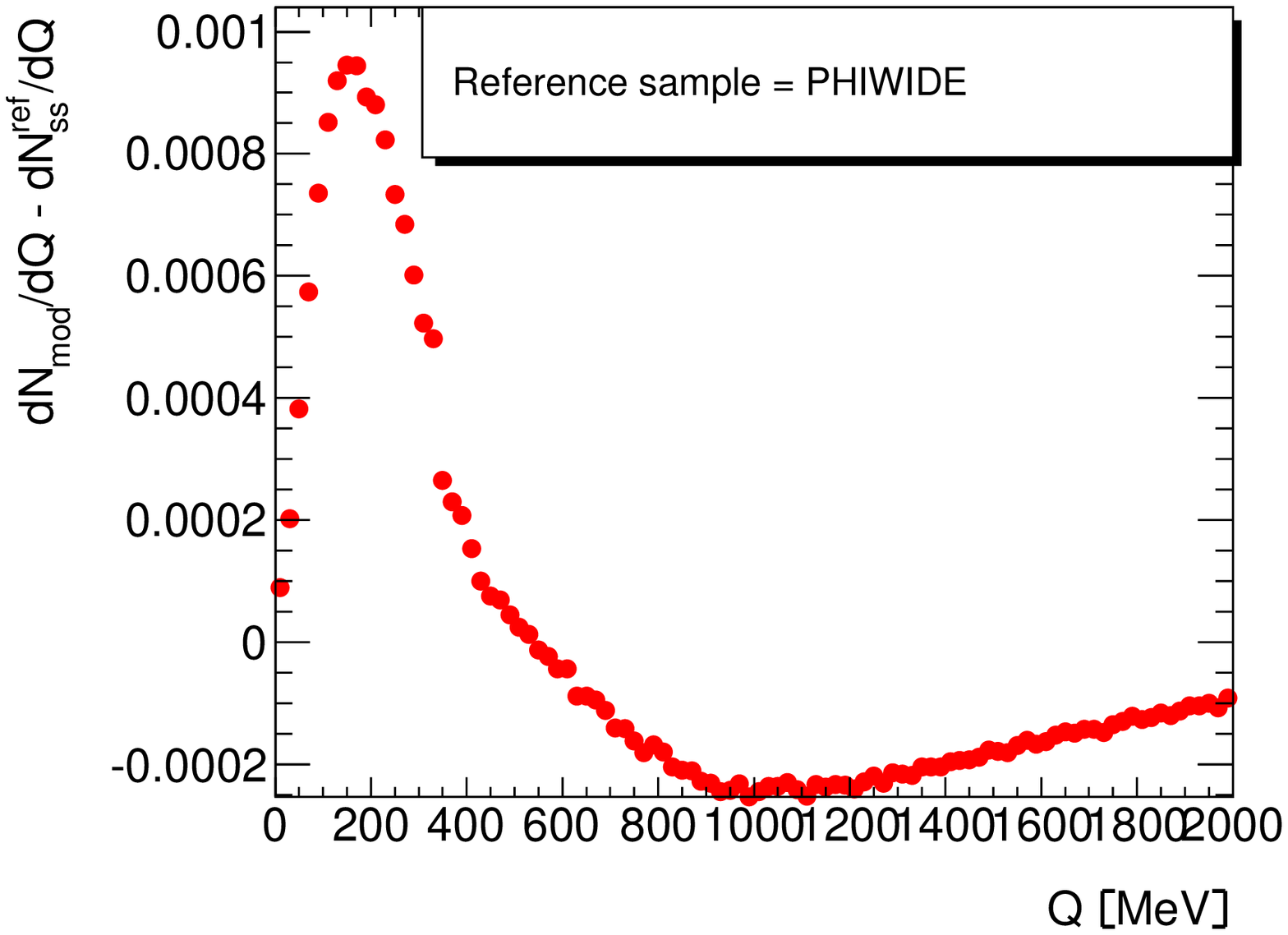}
\label{fig:dif_mod_phiwide}
}
 \subfigure[Fit of the ratio $ R(Q) = \frac{\frac{dN_mod}{dQ}- \frac{dN{_{ss}}^{ref}}{dQ}}{\frac{dN{_{ss}}^{ref}}{dQ}}$ ]{
\includegraphics[width=.5\textwidth]
{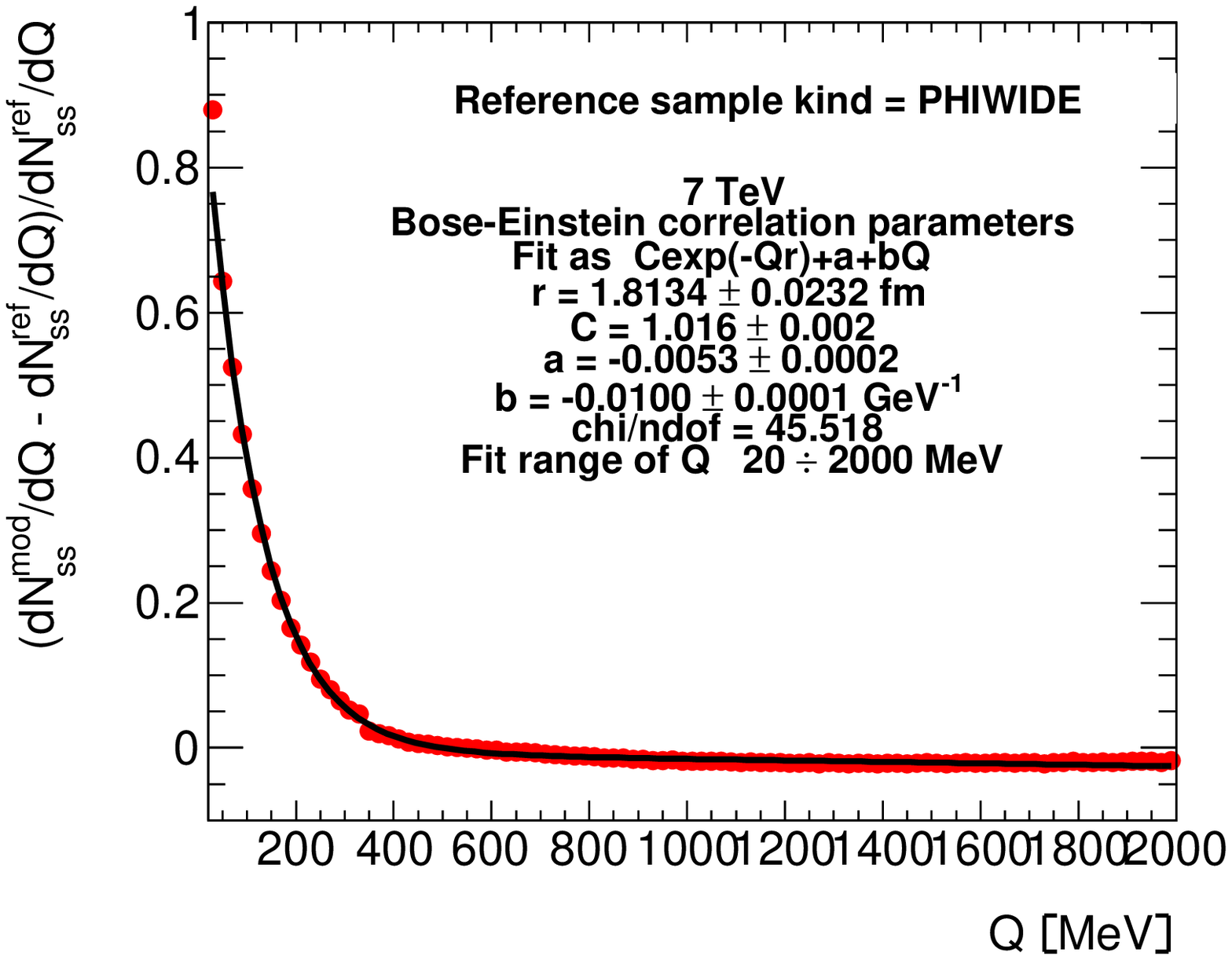}
\label{fig:fit_mod_phiwide}
}
\end{center}
    \caption{ MC Model fit of the ratio R(Q) where the reference sample consists of real events  with all $\vec p_t$
momentum turned by the same random $\delta \phi$  }
     \label{fig:phiwide_case}
\end{figure}

    One gets a bit better reference distribution if  vectors  $\vec p_t$ in an observed event will be turned
    by a random value of
     $\delta \phi$
      \footnote {The "measured" $\phi$ distribution is slightly non uniform because of real detector acceptance and detection efficiencies.
       The same is true for    $\eta$  distribution}
     : the same turn for all tracks - PHIRAND case , or each track is turned by a random angle - PHIWIDE case ( Fig
   ~\ref{fig:phiwide_case}). Fit parameters for the PHIWIDE case is found to be  closer to the model
     than in all previous cases.

\begin{figure}[b]
   \centering
\includegraphics[width=\textwidth]
{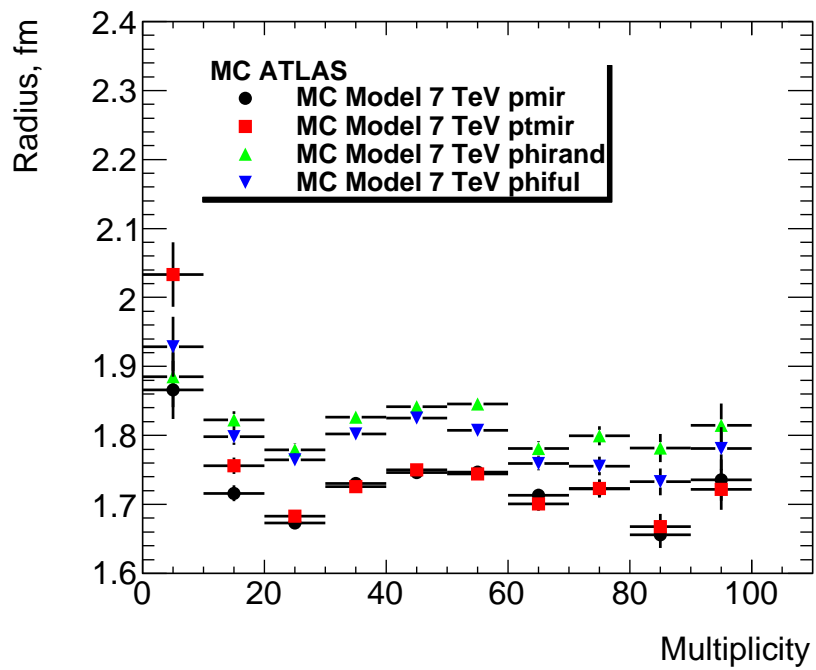}
 \caption{Fit results for different reference samples}
\label{fig:mc_rat_multi}
     \end{figure}

   The model we consider, the  radiation zone radius has a constant value.
    However one might suspect that "imperfection" of reference
   samples is larger at small particles multiplicity. This is the
   case indeed ( Fig ~\ref{fig:mc_rat_multi}). The samples with the reflection algorithms produce quite detectable
   multiplicity
   dependence. The approach with the random $\vec p_t$ turn again looks  better.

   Another important feature is the value of a systematic uncertainty.  One can produce the reference sample of MC events without BEC and
   estimate a value of the source radius with different reference samples. It occurs that the value is quite small
    (~.15 fm, Fig ~\ref{fig:mc_fit_no_BEC})
    and might be consider
    as a contribution of non-BE correlation with small Q value or as   a systematic uncertainty.
    In the following, we will see that  non-BE
    correlation at small Q exists indeed.

\section{2- and 3-dimensional correlations }
   Until now, we have used 1-dimensional distribution - number of events as a function of Lorentz invariant value Q.
This is a natural approach for a simple fit to \emph{measure} phenomenon parameters . Another way have to be
considered in the case of a \emph{search}  of deviations from a phase space prediction with a smooth reaction
amplitude. Usually 2-dimensional correlations is being used.
 Let us consider the traditional correlation plot for like sign pairs  where  differences

  \be
 \begin{split}
         \Delta \phi_{2\pi} = \phi_2 - \phi_1,  \\
         \Delta \eta = \eta_2 - \eta_1
 \end{split}
\ee
\label{dphi2pi}

will be  taken as  arguments - $C_2(\Delta \phi_{2\pi},\Delta \eta)$. Q-value
 and $N_{ch}$ ( number of charged particles in the event) can be used as additional
 arguments to form 3-dimensional correlations
  $C_3(\Delta \phi_{2\pi},\Delta \eta, Q)$ and  $C_3(\Delta \phi_{2\pi},\Delta \eta,N_{ch})$.

\begin{figure}
\begin{center}
 \subfigure[MC $C_2(\Delta \phi ,\Delta \eta)$   for like sign pairs  correlation]{
\includegraphics[width=.5\textwidth]
{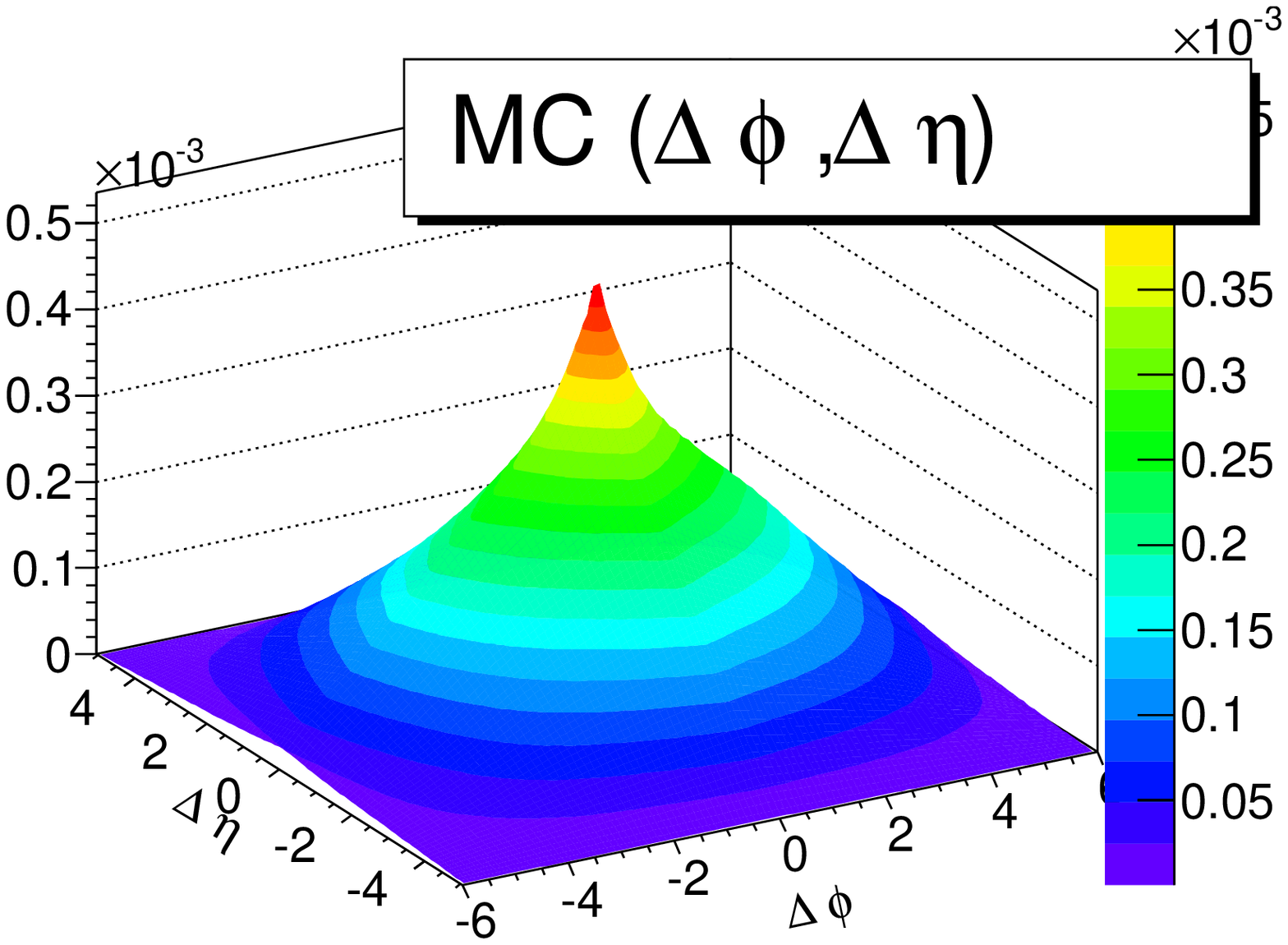}
\label{fig:MC_dphi2pi_deta_like}
} \subfigure[MC $R(\Delta \phi,\Delta \eta)$ =$\frac{C_2(\Delta \phi,\Delta \eta)-C_{2}^{ref}(\Delta \phi,\Delta
\eta)}{C_{2}^{ref}(\Delta \phi,\Delta \eta)}$]{
\includegraphics[width=.5\textwidth]
{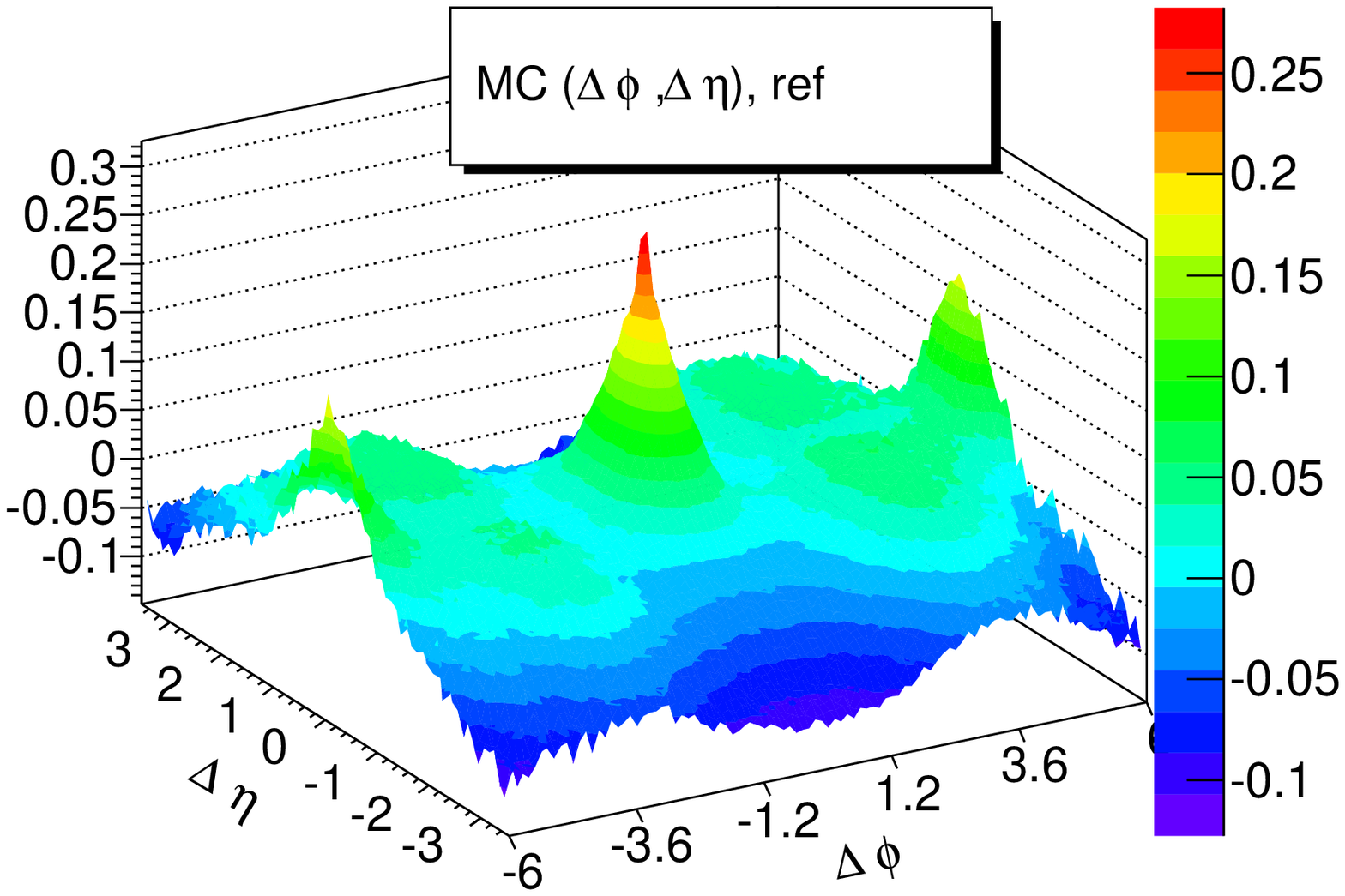}
\label{fig:MC_dif_dphi2pi_deta-like}
}
\end{center}
    \caption{MC $ C_2(\Delta \phi ,\Delta \eta$) for like sign pairs correlation, $ \Delta \phi = \Delta \phi_{2\pi}$  }
     \label{fig:MC_dphi2pi_deta_cor}
\end{figure}

  Fig ~\ref{fig:MC_dphi2pi_deta_cor} shows  several correlation plots.
 If particles would have the uniform distribution for $\eta(-2.5,2.5)$ and for $\phi(-\pi,\pi)$ , then
  the values
$C_1(\Delta \eta )$ and $C_1(\Delta \phi_{2\pi})$ will have triangle distribution in the range $\eta(-5.,5.)$ and
$\phi(-2\pi,2\pi)$. For these reasons  $C_2(\Delta \phi_{2\pi},\Delta \eta)$ looks as a pyramid. Except rather strong
correlation  at $\Delta \phi_{2\pi} \approx \Delta \eta \approx 0$ nothing significant can be seen . One have to
subtract a reference distribution (one of the $p_t$ vector of the pair is turned by a random $\delta \phi$ ) to see
some  additional structure. Unfortunately small nonuniformity might appear artificially because of imperfections of
the reference sample.Some features are connected with definition of  $\Delta \phi_{2\pi}$: evidently,$\Delta
\phi_{2\pi}= \pm 2\pi$ is the same as $\Delta \phi_{2\pi}= 0.$

\begin{figure}
\begin{center}
 \subfigure[MC $R(\Delta \phi,\Delta \eta)$ =$\frac{C_2(\Delta \phi,\Delta \eta)-C_{2}^{ref}(\Delta \phi,\Delta
\eta)}{C_{2}^{ref}(\Delta \phi,\Delta \eta)}$ for like sign pairs]{
\includegraphics[width=.5\textwidth]
{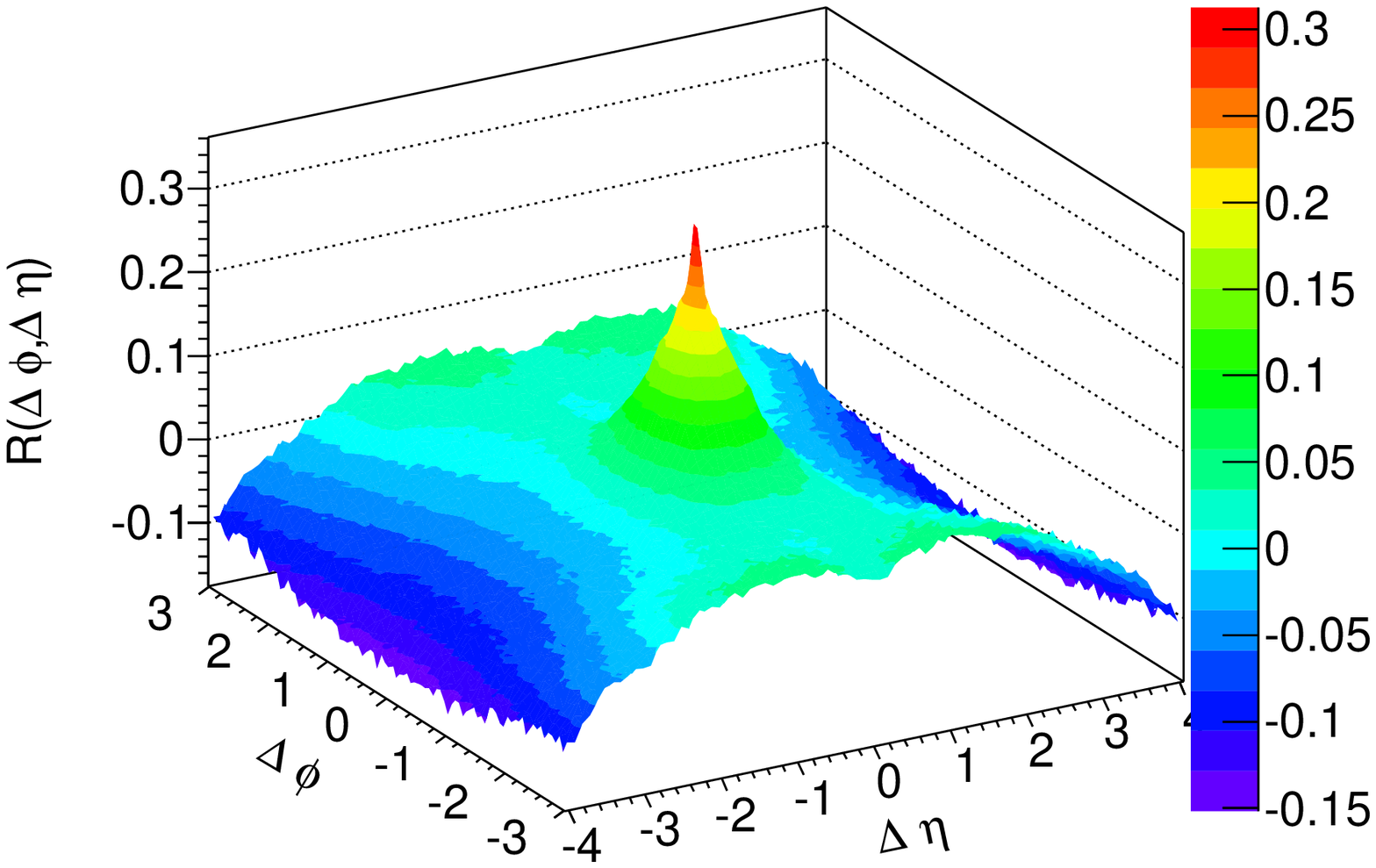}
\label{fig:MC_dif_dphi1pi_deta-like}
} \subfigure[ MC $R(\Delta \phi,\Delta \eta)$ =$\frac{C_2(\Delta \phi,\Delta \eta)-C_{2}^{ref}(\Delta \phi,\Delta
\eta)}{C_{2}^{ref}(\Delta \phi,\Delta \eta)}$ for unlike sign pairs]{
\includegraphics[width=.5\textwidth]
{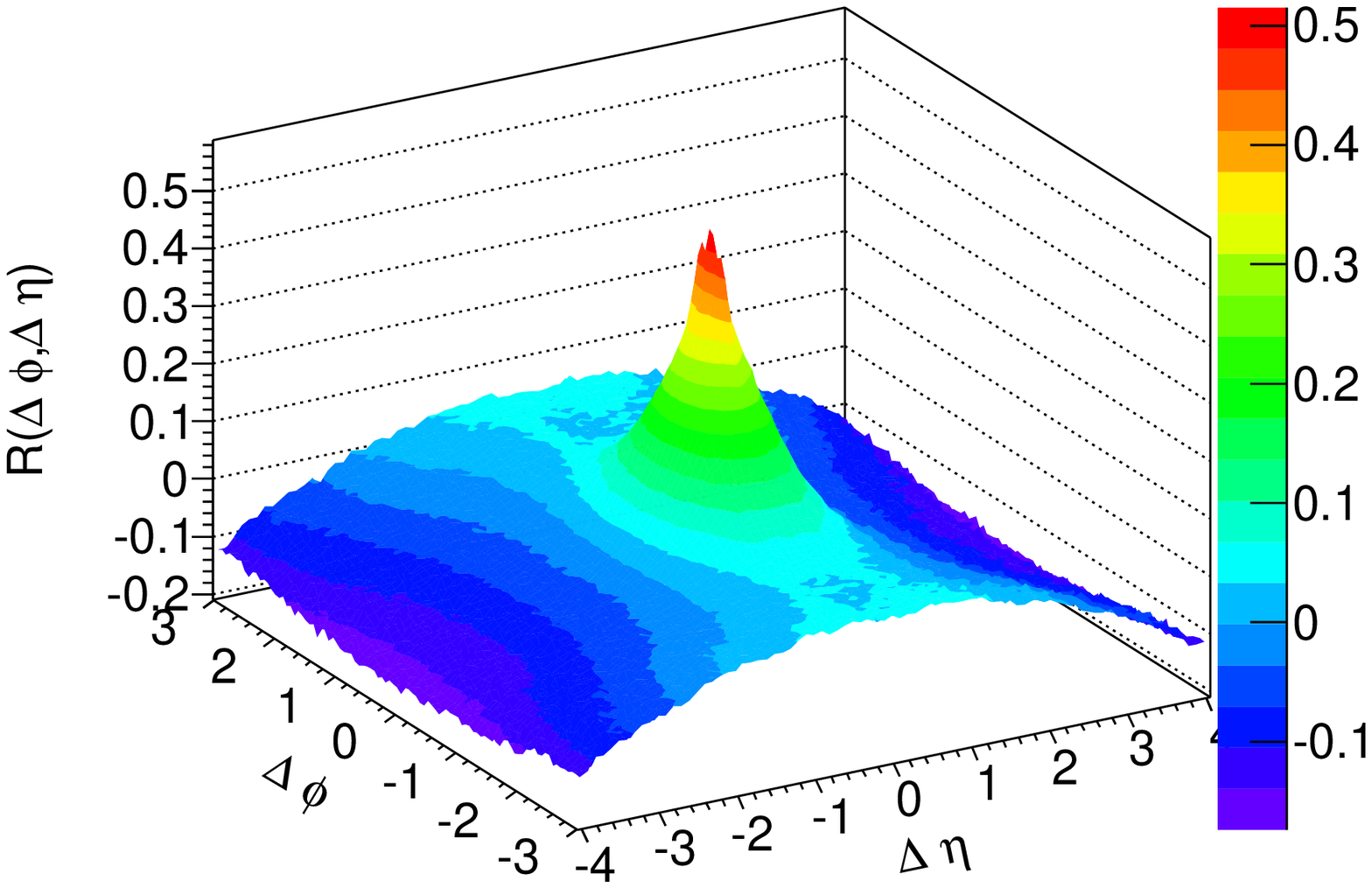}
\label{fig:MC_dif_dphi1pi_deta-unlike}
}
\end{center}
    \caption{MC $C_2( \Delta \phi ,\Delta \eta$) ,$\Delta \phi = \Delta \phi_{\pi} $ }
     \label{fig:MC_dphi1pi_deta_cor}
\end{figure}

It is better    to change the definition of $\Delta \phi$ :
 \be
 \begin{split}
      \Delta \phi_{\pi} = \phi_2 - \phi_1,  |\phi_2 - \phi_1| < \pi \\
      \Delta \phi_{\pi} = \phi_1 - \phi_2,  |\phi_2 - \phi_1| > \pi
\end{split}
\ee
\label{dphi1pi}
   $\Delta \phi_{\pi}$ will have  a uniform distribution in the range ($
-\pi,\pi$) in the absence of correlations and to study  correlations  one has no need for any reference to subtract.
Fig ~\ref{fig:MC_dphi1pi_deta_cor} shows $C_2( \Delta \phi, \Delta \eta)$ before and after reference subtraction.
Similar plot for unlike sign pairs is shown also . Strong correlation is independent on charge of particles.

\begin{figure}
\begin{center}
 \subfigure[MC $ C_2(Q,\Delta \eta $) correlation]{
\includegraphics[width=.5\textwidth]
{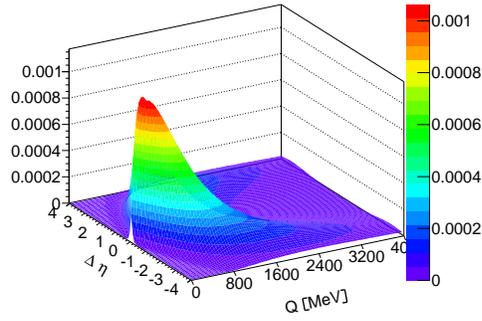}
\label{fig:MC_Q_deta}
}
 \subfigure[ MC $C_2(Q, \Delta \phi $) unlike sign correlation]{
\includegraphics[width=.5\textwidth]
{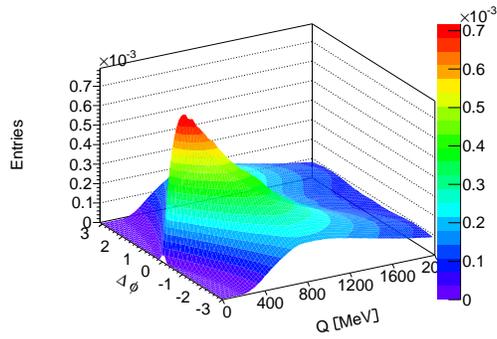}
\label{fig:MC_dphi_like-unlike}
}
\end{center}
    \caption{ MC $ C_2(Q,\Delta \phi)$ and $C_2(Q,\Delta \eta $) correlations }
     \label{fig:MC_Q_deta_dphi}
\end{figure}

\begin{figure}[b]
   \centering
\includegraphics[width=0.8\textwidth]{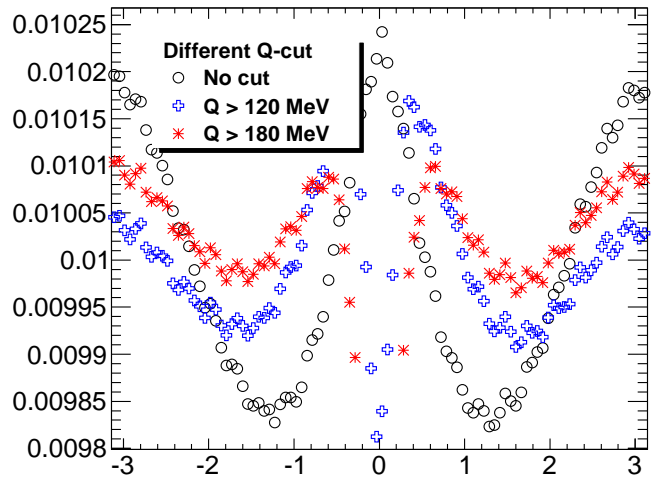}
\caption{ MC $C_1(\Delta \phi$) distribution with different cuts in Q value }
\label{fig:MC_dphi_q_cut}
     \end{figure}

 Let us consider other correlations $C_2( \Delta \phi ,Q)$ and
 $C_2(\Delta \eta ,Q)$ (Fig ~\ref{fig:MC_Q_deta_dphi}).
An enhancement at small Q value can be expected. Q is  small if $ \Delta \phi \sim0$ and $ \Delta \eta \sim0$ (
momentum value of particles in pairs are larger than masses) ,\emph{independent on the momentum values and the
charges of particles}. At the same time at large Q value there is a dip at  $ \Delta \phi \sim0$
 and ( $ \Delta \eta\sim0$ ).
 Then after integration in Q , there might be full compensation of the enhancement at small Q , if \emph{there is no dynamical reason
 to make an enforcement of this  enhancement}. By other words, with an appropriate cut in
 Q value, small $ \Delta \phi$ and $ \Delta \eta$ correlation can be suppressed (Fig ~\ref{fig:MC_dphi_q_cut}).
  An effect of dynamical correlations exists only when Q is less than $\sim 200 MeV$. After such cut nonuniformity
  in $\Delta \phi$ decreases from  $\sim 4\%$ to $\sim 1\%$.
  Very important that such behavior is the same for like sign and unlike sign pairs
  ( except natural traces of resonances at $Q>400$ MeV) .

\begin{figure}
\begin{center}
   \subfigure[ MC $R(\Delta \phi,\Delta \eta)$  correlation without Q cut]{
 \includegraphics[width=.5\textwidth]
{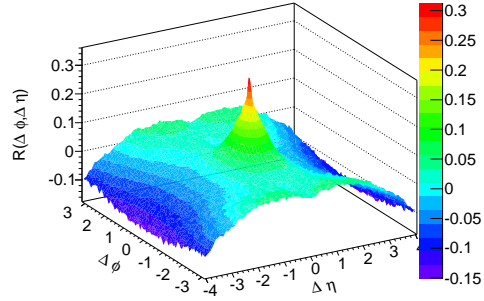}
\label{fig:MC_rat2_nocut}
}
 \subfigure[ MC $R(\Delta \phi,\Delta \eta)$  correlation with $Q > 120$ MeV cut]{
\includegraphics[width=.5\textwidth]
{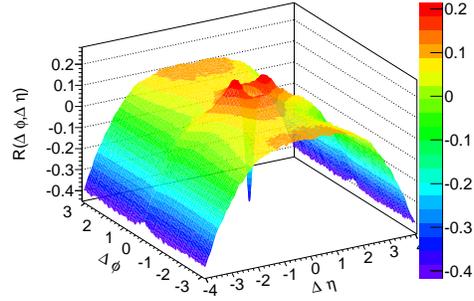}
\label{fig:MC_rat2_120cut}
}
 \subfigure[MC $R(\Delta \phi,\Delta \eta)$  correlation with $Q > 200$ MeV cut]{
\includegraphics[width=.5\textwidth]
{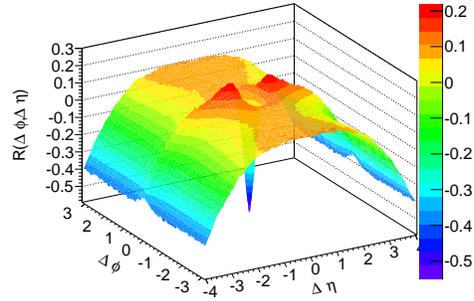}
\label{fig:MC_rat2_200cut}
}
\end{center}
    \caption{MC $R(\Delta \phi,\Delta \eta)$  correlation }
     \label{fig:MC_deta_dphi_q_cut}
\end{figure}

\begin{figure}
\begin{center}
   \subfigure[ MC $R(\Delta \phi,\Delta \eta)$  correlation without Q cut]{
 \includegraphics[width=.5\textwidth]
{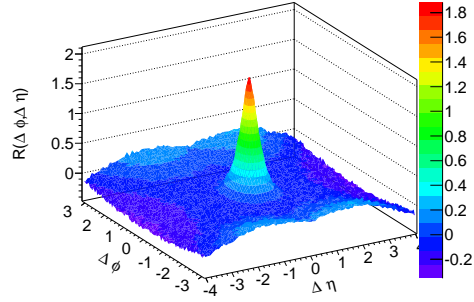}
\label{fig:MC_p1GeV_rat2_nocut}
}
 \subfigure[ MC $R(\Delta \phi,\Delta \eta)$  correlation with $Q > 120$ MeV cut]{
\includegraphics[width=.5\textwidth]
{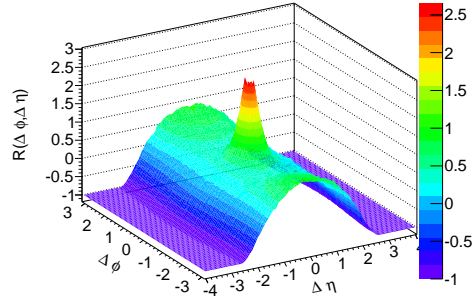}
\label{fig:MC_p1GeV_rat2_120cut}
}
 \subfigure[ MC $R(\Delta \phi,\Delta \eta)$  correlation with $Q > 200$ MeV cut]{
\includegraphics[width=.5\textwidth]
{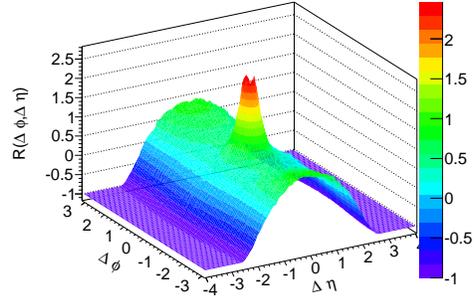}
\label{fig:MC_p1GeV_rat2_200cut}
}
\end{center}
    \caption{MC $R(\Delta \phi,\Delta \eta)$  correlation for pairs with $p_t^{min} = 1 GeV$ }
     \label{fig:MC_p1GeV_deta_dphi_q_cut}
\end{figure}

    The next step is to have a look an influence of such Q cuts on 2-dimensional plots (Fig
    ~\ref{fig:MC_deta_dphi_q_cut}).  The "Long-Range, Near-Side Correlations" (ridge) starts to be appear. Such
    effect has not been seen in MC before ( see \cite{CMS-RIDGE})not only because Q-cut never was used before, but also because
    of different way to construct the reference sample ( see in the follows). The ridge was observed  \cite{CMS-RIDGE}
    in the data analysis of    pp-collisions at the LHC at
    7TeV in high multiplicity events with selection of pairs of charged particles with $p_t$ in the range (1-3) GeV.
    In MC distribution there is no trace of a ridge, if only small $p_t$ cut has been done.
    Fig  ~\ref{fig:MC_p1GeV_deta_dphi_q_cut} shows the correlations for particles with $p_t > 1 GeV$.
   The multiplicity cut only decrease statistics but does not change the conclusion. By other words, the selection of
   pairs of particles with $Q > 200$ MeV can be good tool for search of a ridge effect.

 \section{Conclusions}
Rather simple MC model for  Bose-Einstein Correlations is used to judge the quality of different ways to extract the
BEC from observable distribution. Several emulated reference samples have been considered. The reference sample with
random turn of observed transverse momentum vector $\vec p_t$ ( $\phi$ randomization ) provides the fit results
closest to the model values.

  Such approach is implemented for the analysis of $C_2(\Delta \phi, \Delta \eta$) correlation. In the  traditional
  definition of $\Delta \phi_{2\pi}$, to extract correlations the subtraction of a reference distribution is the must.
  New definition $\Delta \phi_{\pi}$ as the angle between $\vec p_{t1}$ and  $\vec p_{t2}$ makes $\Delta \phi_{\pi}$ distribution
  uniform  with small contribution from an intrinsic correlations.
  For this reason the $C_1(\Delta \phi_{\pi})$ can be studied without any reference.
   It is shown that at
  small Q-value there might be  two kind of $C_1(\Delta \phi_{\pi})$ correlation - kinematical and dynamical. At large Q there is
  $C_1(\Delta \phi_{\pi})$ \emph{anticorrelation}. Then so called short-range correlation ( $\Delta \phi_{\pi} \sim 0$ )
   can be suppressed  by an appropriate cut  in the Q- value.

\newpage

\begin{figure}
\begin{center}
 \subfigure[The  difference of normalized distributions  $ \frac{dN_{mod}}{dQ}-\frac{dN{_{ss}}^{ref}}{dQ}$ ]{
\includegraphics[width=.5\textwidth]
{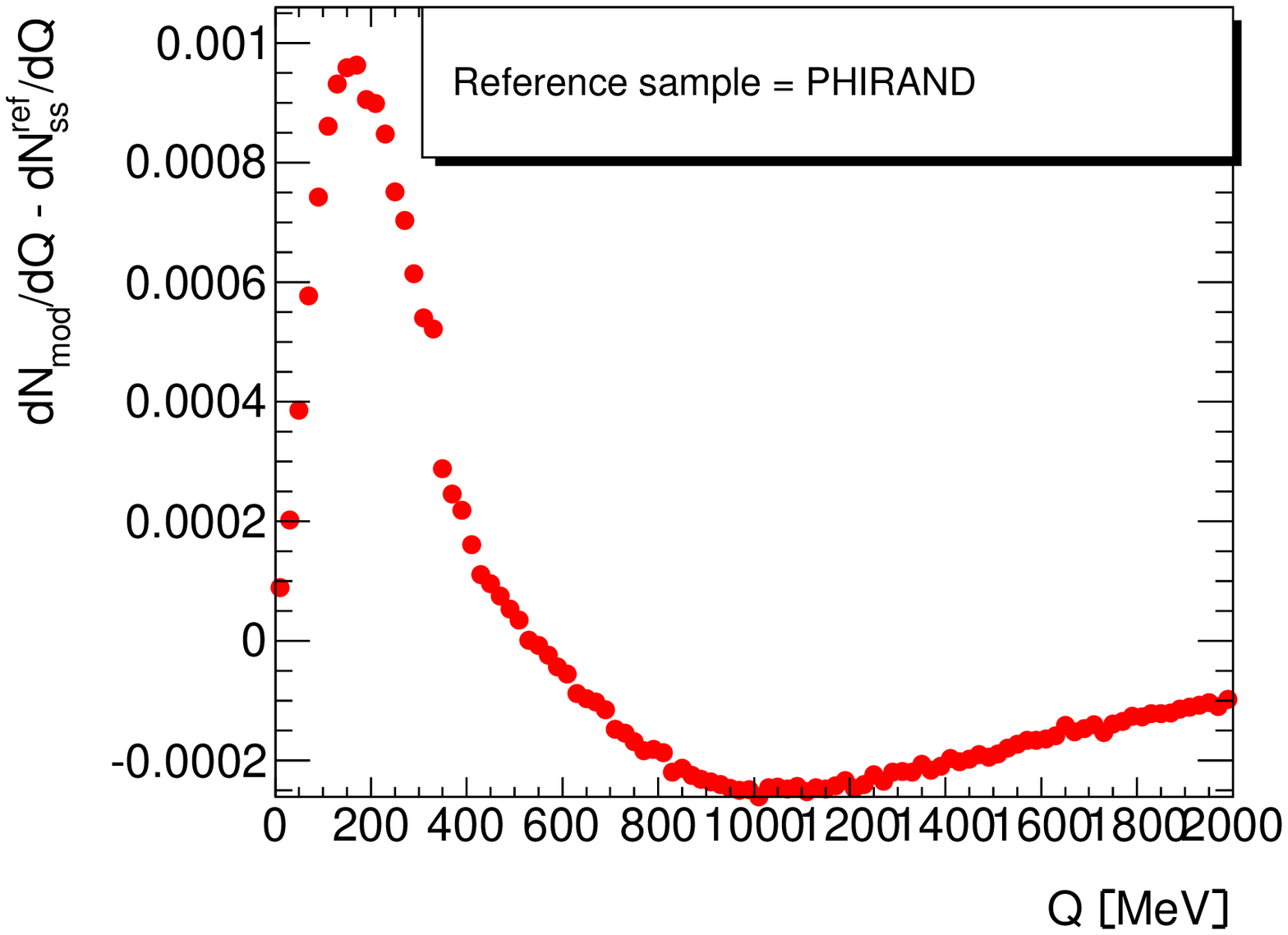}
\label{fig:dif_mod_phirand}
}
 \subfigure[Fit of the ratio $ R(Q) = \frac{\frac{dN_mod}{dQ}- \frac{dN{_{ss}}^{ref}}{dQ}}{\frac{dN{_{ss}}^{ref}}{dQ}}$ ]{
\includegraphics[width=.5\textwidth]
{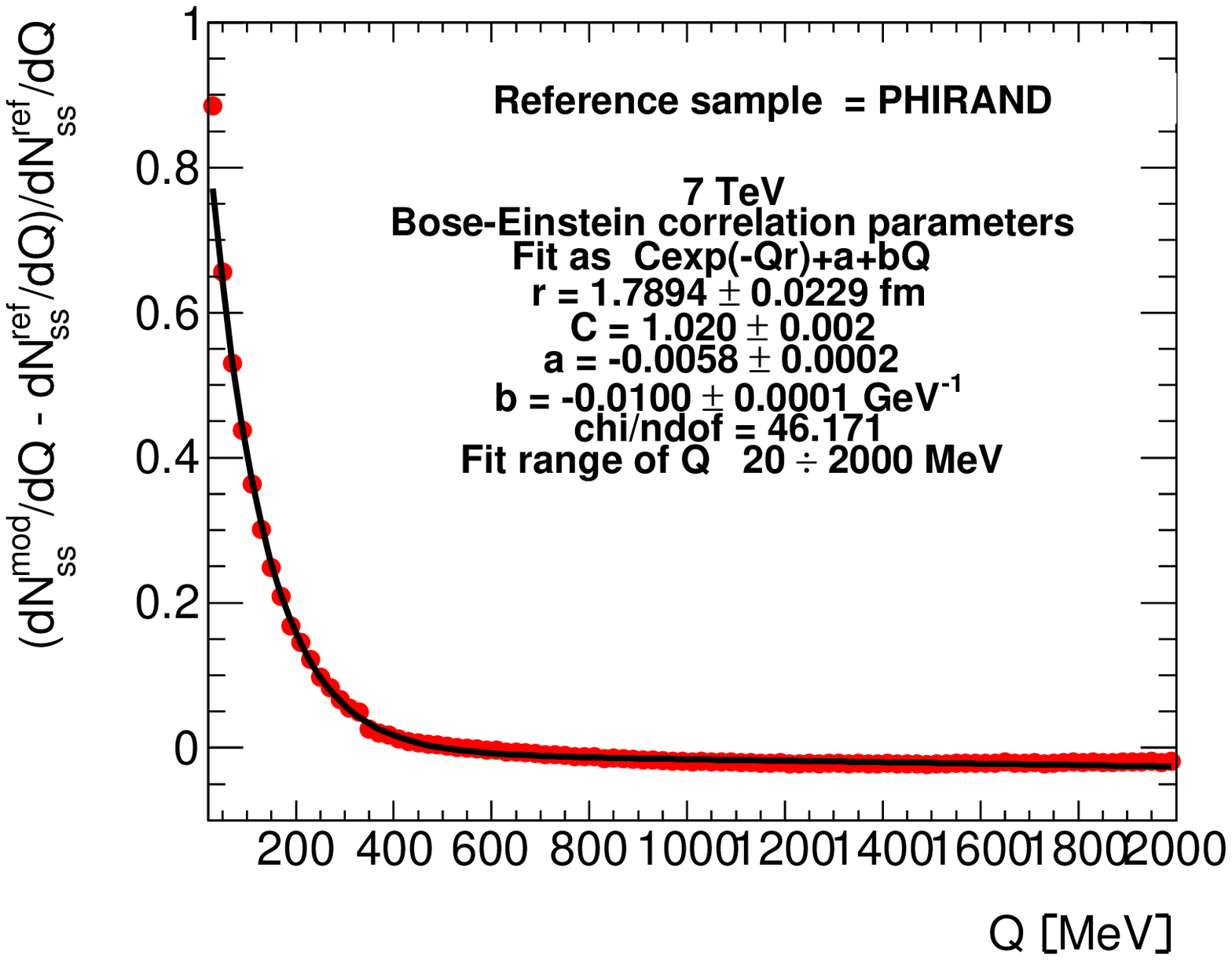}
\label{fig:fit_mod_phirand}
}
\end{center}
    \caption{ MC Model fit of the ratio R(Q) where the reference sample consists of real events  with all $\vec p_t$
momentum turned by the same random $\delta \phi$  }
     \label{fig:phirand_case}
\end{figure}

\end{document}